\documentclass[twocolumn,showpacs,aps]{revtex4}
\usepackage{graphicx}
\usepackage{dcolumn}
\begin{document}
\title{Novel Electron Spectroscopy of Tenuously and Weakly Bound Negative Ions}
\author{A.\ Z.\ Msezane}
\affiliation{Center for Theoretical Studies of
        Physical Systems, Clark Atlanta University, Atlanta, Georgia 30314 USA}
\author{Z.\ Felfli}
\affiliation{Center for Theoretical Studies of
        Physical Systems, Clark Atlanta University, Atlanta, Georgia 30314 USA}
\author{D.\ Sokolovski}
\affiliation{School of Mathematics and Physics,
     Queen's University of Belfast,
                  Belfast, BT7 \#1NN, United Kingdom}
                  \date{\today}
\abstract
A novel method is proposed that uses very slow electron elastic collisions with 
atoms to identify their presence through the observation of tenuously bound 
(electron impact energy, $E<0.1 eV$) and weakly bound ($E<1 eV$) negative ions, 
formed as Regge resonances during the collisions.}
\maketitle
The crucial importance of the lanthanide atoms in general and in the design and 
synthesis of novel functional compounds has recently stimulated considerable 
research in the context of superconductivity\cite{1,2,3}, Bose-Einstein 
condensation\cite{4} and photoluminescence\cite{5}.  Specifically, 
photoluminescence has been observed for Sr$_2$CeO$_4$ through doping with the ions of 
Eu, Sm or Yb\cite{5}, a Bose-Einstein condensation (BEC) of Yb atoms has been 
realized by evaporative cooling, including the continuing search for BEC of 
atoms with no electron spin in two electron systems such as the alkaline-earth 
and Yb atoms\cite{4} and the fabrication of co-doped SrFe$_2$As$_2$ thin films on 
(La,Sr)(Al,Ta)O$_3$(001) has been accomplished\cite{1}. The utility of the lanthanide 
atoms in heavy fermion metals and heavy fermion compounds, whose most attractive 
and elusive phenomenon is the appearance of superconductivity in a few of them, 
requires a thorough investigation and understanding of the structure and dynamics 
of very low-energy electron elastic scattering.  The presence of the partly filled 
f-orbitals causes them to behave like localized magnetic moments\cite{6}.  
Furthermore, the rare-earth monosulfides like LaS, ErS, EuS, GdS, {\it etc.} at various 
III-V semiconductor surfaces offer the attractive possibility of reaching negative 
electron affinities\cite{7}. 

Here we propose the spectroscopy of very slow $E<1eV$ electrons realized through 
elastic collisions with various atoms, resulting in electron attachment to form 
temporary negative ions as Regge resonances.  Unique dramatically sharp long-lived 
resonances in the total elastic cross sections (TCS's) have recently been 
identified as signatures of the stable bound states of the negative ions formed 
during the collisions between the slow incident electrons and the target neutral 
atoms\cite{8,9,10,11}.  The imaginary part of the complex angular momentum L, 
Im L, has been used to distinguish between the resultant stable bound states 
of the negative ions and the shape resonances; Im L for the 
former is several orders-of-magnitude smaller than that for the latter.  
The investigations used the recently developed Regge-pole methodology \cite{12} 
together with a Thomas-Fermi type potential that incorporates the vital 
core-polarization effect to calculate the near-threshold electron elastic TCS's 
and the Mulholland partial cross sections (MPCS's) as well as to obtain the 
Ramsauer-Townsend (RT) minima, the shape resonances and the Wigner threshold 
law.  The proposed low-energy spectroscopy will be useful to uniquely identify 
the presence of various atoms in different environments as well as to determine 
the appropriate combinations of atoms for exotic molecular/materials designs. 

Understanding the structure and the dynamics of low-energy electron elastic 
collisions, resulting in the formation of negative ions as Regge resonances, 
is quite challenging and the measurements of absolute cross sections for the 
process are difficult to obtain.   Temporary negative ion states and their 
properties, represented essentially by shape resonances, are responsible for 
the mechanism through which low-energy electron scattering deposits energy and 
induces chemical transitions\cite{13}.  Electron-induced chemical processes 
resulting in negative ion production have been interpreted in terms of shape 
resonances\cite{14}.  The Wigner threshold law\cite{15} is essential in high 
precision measurements of binding energies (BE's) using photodetachment 
threshold spectroscopy\cite{16}.  Atoms with small electron affinities (EA's) 
(EA corresponds to the energy with which the additional electron is bound to 
the neutral atom; it is numerically equal to the BE) have been proposed for 
the quenching of Rydberg atoms through collisions\cite{17} as well as in Dry 
Pile\cite{18}. The selection of molecules for cold and ultracold experiments, 
including their creation\cite{19}, requires knowledge of the spin-flipping 
and elastic cross sections\cite{20};  the former can also affect buffer-gas 
loading of ultra-cold molecules.  The RT effect is essential in understanding 
{\it inter alia} sympathetic cooling and production of cold molecules from 
natural fermions\cite{21}.  

The fundamental understanding of the near-threshold electron scattering process, 
with attendant electron attachment, is also essential in and will benefit the 
investigation of the formation of BEC of alkali-earth atoms and the utility of 
quantum degenerate Yb atoms in fundamental physics research for optical clocks,
qubit arrays\cite{22} and parity violation\cite{23}.  The investigation of the 
EA's  and ionization potentials of the 4d and 5d transition metal elements
by density functional methods\cite{24}, will also benefit significantly through 
reliable EA's.  We note that our methodology extracts the EA values without 
{\it a priori} knowledge of them. Importantly, the present data are also expected 
to aid the planned investigation and understanding of superconductivity in the 
compounds La$_2$CdSe$_2$O$_2$, LaMnP{\it n}O and LaCuSeO\cite{1}. 

The Mulholland formula\cite{25}, implemented within the complex angular momentum 
representation of scattering\cite{12, 26} given by Eq. (8)\cite{12} has been
employed for the calculations of the TCS's.  All the cross sections presented 
in Fig.1 were computed using this equation, including those for the lanthanides
\cite{27}.  The details of the numerical calculations are found in\cite{12, 27} 
and atomic units are used throughout the paper.  For the scattering potential 
we used the well investigated form of the TF potential\cite{28} by Belov 
{\it et al}\cite{29} given by Eq. (2)\cite{11}.  For small radial distances, 
the potential describes the Coulomb attraction between an electron and a nucleus, 
while at large distances it mimics the polarization potential and accounts 
properly for the vital polarization interaction at very low energies.
In the Regge pole methodology the Schrodinger equation  is solved for complex 
L values using positive E values.  Im L is then used to distinguish between the 
shape resonances (short-lived resonances) and the stable bound states of the 
negative ions (long-lived resonances) formed as Regge resonances in the electron-
atom scattering.

The region $ E < 0.10eV$ corresponds to the energy region of the tenuously 
bound negative ions formed during the collisions. Here the TCS's, with the 
characteristic sharp peaks, for electron scattering from some atoms, particularly 
the lanthanides are displayed, as shown in Fig. 1 (the detailed cross sections 
for all the lanthanides are found in\cite{27}).  In the ln x ln representation  
of the TCS's versus E (eV) the various sharp resonances, whose positions are 
identified with the BE's of the negative ions formed during the collisions, are 
emphasized.  In the limit $ E \rightarrow 0$ starting from 0.1 eV the electron 
attachment cross sections, represented by the various peaks in Fig. 1 have 
interesting features.  Two limiting curves determine the TCS's behavior: one is 
associated with the variation of the e$^-$-Ca attachment cross section with 
Re L=1 (red top curve) attachment and a Wigner threshold law, determined by 
an n=6 MPCS (see Fig. 1\cite{8}).   The other whose Wigner threshold law is 
defined by the n=7 MPCS (green curve) corresponds to electron attachment to 
the atoms Tm, Hf, Yb, Lu and Sr, 
labeled accordingly in the figure. These two limiting curves are very 
important in the context of the Wigner threshold law for negative ions in general 
and for guiding measurements in particular. 

\begin{figure}
\includegraphics{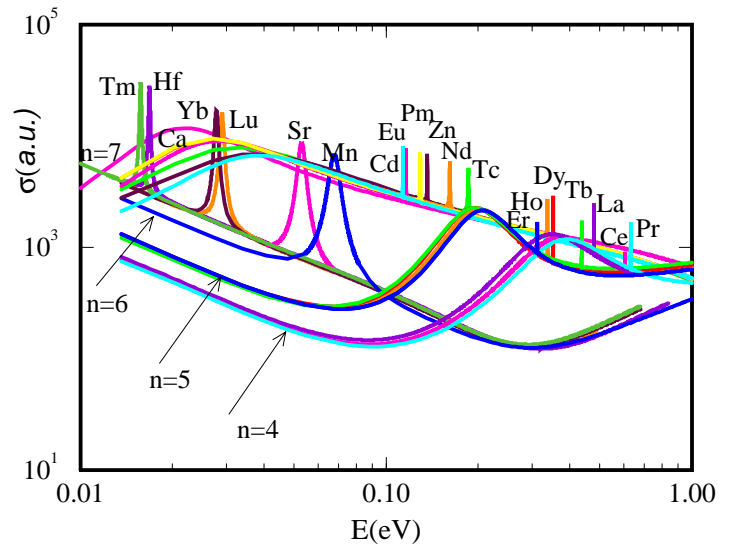}
\caption{
TCS's ({\it a.u.}) for low-energy electron scattering by Tm, Hf, Yb, Lu, Ca, 
Sr and Mn atoms in the energy region of the formation of their tenuously bound 
negative ions, with Re L=2, except for Ca whose Re L=1; 
they follow a Wigner threshold law that is determined by the n=7 MPCS. The atoms 
Cd, Eu, Pm, Zn, Nd and Tc form weakly bound negative ions with Re L=3, while 
Er, Ho, Dy, Tb, La, Ce and Pr form similar ions as well but with Re L=4.  The 
many thin films\cite{1} are pivoted by the La and Sr atoms; their importance is 
clearly manifested in the figure.  Note also the positions of Mn and Cd in the 
figure and their roles in the compounds La$_2$CdSe$_2$O$_2$ and LaMnP{\it n}O\cite{1}.}
\end{figure}

The region $0.1 < E < 1.0 eV$, defining the formation of weakly bound negative 
ions, can be divided into several segments.  One of the most interesting of 
these is the energy region between 0.1 and 0.2 eV.  The peaks here correspond 
to electron attachment to Re L=3 and the energy range reflects the range of BE's 
possible in the attachment process. The behavior of the TCS's of these atoms 
in this region can be understood through the scrutiny of the typical example 
of e$^-$-Eu scattering, shown in Fig. 2.  Except for the presence of the very 
sharp peak at 0.116 eV (whose position defines the BE of the resultant Eu$^-$ 
negative ion) the TCS resembles essentially that for the e$^-$-Ca scattering, 
see Fig. 1\cite{8}.  The near-threshold behavior of the e$^-$-Eu 
scattering also resembles that for the e$^-$-Ca scattering; however, the unique 
sharp peak at 0.116 eV on the n=5 MPCS corresponds to the electron attachment cross 
section during the collision. Notably, the dominant shape resonance in Fig. 2 is 
suppressed in the representation of Fig. 1, thereby revealing a clear dominance 
of the attachment cross section.  All the TCS's for the atoms shown in this energy 
region are typified by the e$^-$-Eu cross sections of Fig. 2.  We note the 
presence of maxima around E=0.2 eV and these will be discussed in another section 
below. In this energy range the electron attachment is determined by Re L=3 while 
the n=5 MPCS determines the Wigner threshold law, see Fig. 2.  These results 
demonstrate the importance of the representation of the cross sections as in 
Fig. 1 and are an important check of the measured or calculated results without 
invoking Im L\cite{27}.  

\begin{figure}
\includegraphics{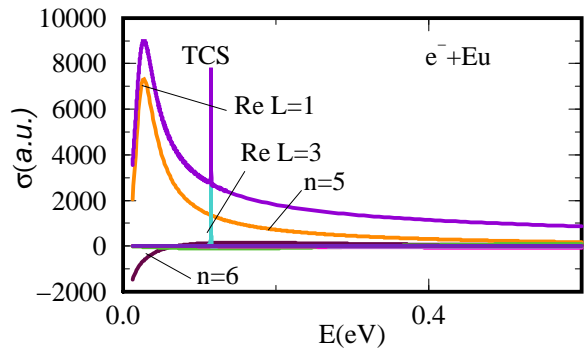}
\caption{
TCS and MPCS's ({\it a.u.}) for e$^-$-Eu scattering versus E (eV); note the 
pronounced maximum near threshold, which is dominated by a MPCS corresponding 
to Re L=1, a shape resonance at 0.029 eV.  The Regge trajectory passing 
near Re L=3 at E=0.116 eV determines the bound state of Eu$^-$.}
\end{figure}

As the energy increases (Fig. 1) from 0.3 eV in the region $0.3 eV< E < 0.45 eV$ 
the various resonances, whose positions correspond to the relevant BE's, begin 
to overlap.  Let us focus on the resonances corresponding to electron attachment 
to the atoms Er, Ho, Dy, and Tb; they all form stable weakly bound negative ions 
with electron attachment to Re L=4 and BE's between 0.3 and 0.45 eV (see also\cite
{27}).  Their TCS's and MPCS's are typified by that for the e$^-$-Dy scattering, 
shown in Fig. 3.  The attachment resonances for these atoms are preceded by their 
shape resonances at around 0.2 eV and RT minima at about 0.08 eV. The maxima at 
around 0.2 eV correspond to the positions of the shape resonances of Er, Ho, Dy 
and Tb.  At about 0.3 eV the minima of the cross sections for electron 
scattering by Tm, Hf, Yb, Lu, Sr and Mn, with electron attachment of Re L=2 
appear.  At this energy the observation of the various resonances corresponding 
to electron attachment to Re L=4 is facilitated, including the identification 
of the shape resonances of the Ce, La and Pr atoms. 

\begin{figure}
\includegraphics{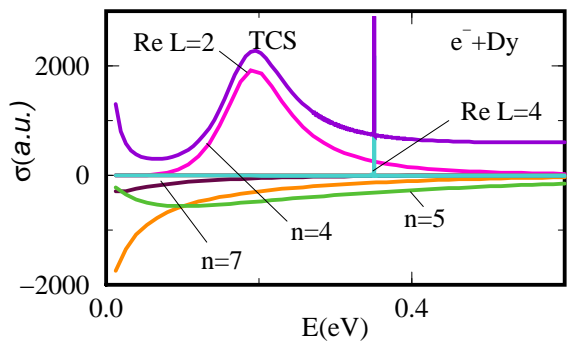}
\caption{
The same as in Fig. 2 except that the results are for e$^-$-Dy scattering. 
The Regge trajectory passing near Re L=2 at E=0.194 eV defines a shape 
resonance, while that passes near Re L=4 at E=0.3515 eV determines the bound 
state of Dy$^-$.  Note the RT minimum at 0.068 eV and the Wigner threshold 
behavior of the TCS, determined by the n=5 trajectory (see also Fig. 1).
}
\end{figure}

In Fig. 1 the variation of the TCS's with E for the La, Ce and Pr atoms resembles 
that for the Er, Ho, Dy and Tb atoms, except that they rest on a trajectory 
which eventually becomes a n=4 MPCS as E $\rightarrow$ 0.  The electron attachment 
cross sections for these atoms are also determined by Re L=4 and have shape 
resonances at around 0.35 eV\cite{27}.  Here the attachment energy is at a higher 
value compared to those in the previous set.  For the lanthanides, as the Z 
increases the interaction potential increases correspondingly and pulls the 
positions of the resonances closer to threshold.  We note that the Wigner threshold 
behavior of this set is determined by the n=4 MPCS rather than the n=5 of the 
previous set.  

In summary, the utility and strength of the representation of the elastic TCS's 
for various combinations of atoms have been demonstrated for possible novel 
materials design and fabrication as well as quenching.  This could also be 
useful for identifying the presence of various atoms in gaseous mixtures through 
the observation of the formation of tenuously bound and weakly bound negative 
ion states.  Figure 1 could also be placed on a microchip for convenient detection 
and check of measured or calculated EA's.  The most significant revelation of the 
representation of the TCS's versus E in Fig. 1 is that near the intersection of 
the limiting curves in the region $E \rightarrow 0.01$ eV, an assumed Re 
L=1\cite{30} or Re L=2 electron attachment yields an EA that is close to the 
correct value.

Work supported by U. S. DOE, Division of Chemical Sciences, Office of Basic 
Energy Sciences, Office of Energy Research and the NSF funded CAU CFNM.
The computing facilities at the Queen's University of Belfast, UK and of DOE 
Office of Science, NERSC are reatly appreciated.

{}
\end{document}